# Uncertainty Analysis in 3D SPECT Reconstruction based on Probabilistic Programming

Manu Francis[1], Muhammed Tarek[1], Mark Pickering[1] and Murat Tahtali[1]

[1]SEIT, UNSW Canberra, Australia
E-mail: manu.francis@student.adfa.edu.au

**Abstract**

Single Photon Emission Computed Tomography (SPECT) is one of the nuclear medicine imaging modalities used for functional analysis of animal and human organs. Gamma rays emitted from the scanned body are filtered with collimators and detected by the SPECT head that is composed of an array of gamma detectors. The conventional reconstruction algorithms do not deem the uncertainty level that is associated with the field of view of SPECT collimators. In this paper, we incorporate the probabilistic programming approach for SPECT image reconstruction. No-U-Turn Sampler (NUTS) is used to estimate the scanned object system by considering uncertainty. The obtained results indicate that the presented work in 3D SPECT image reconstruction surpassed over the conventional reconstruction methods in terms of generating uncertainty information. However, the reconstruction time needs to be improved further for phantom sizes of 128x128x128 voxels and higher.

Keywords: Image Reconstruction, Probabilistic Programming, Uncertainty Analysis, SPECT

## 1. Introduction

Single Photon Emission Computed Tomography (SPECT) systems use emitted gamma radiation. This gamma radiation is generated by the infused radiolabelled medicines into the living objects that need to be scanned. The emitted gamma radiation is filtered through collimators and detected by gamma detectors made up of a scintillator and an array of photomultipliers. Projection images represent the distribution of the radionuclide from 3D to a 2D plane for a specific view angle [1]. The projection for one view angle is not sufficient to estimate the distribution of the gamma sources inside the object. Hence, multiple projection images at different view angles are obtained to reconstruct the gamma emission distribution. These multiple projections are generated by rotating the SPECT gamma camera at discrete angles around the scanned body.

The mathematical representation that models the forward projection SPECT imaging is:

$$g = Af \quad (1)$$

Where $g$ is the projection, $A$ is the system matrix, and $f$ is the scanned phantom. The image reconstruction finds $f$ by solving Equation 1.

The distribution of the radionuclide can be identified from the projection images by using certain reconstruction algorithms. The main idea behind the reconstruction relies on mapping the radionuclide distribution by using the back-projection operator and projection images, as discussed in next section. Analytical and iterative methods are the two main reconstruction algorithm categories.

Analytical methods are simple and mainly based on back-projection. In these methods, the scanned object is reconstructed slice by slice. The projection of each sample is arranged row-wise, called a sinogram is used in the reconstruction algorithm. Each element is back-projected to the area of interest, based on the direction of the radiation. The shape and orientation of the collimator provide information about the direction of the radiation. Each voxel in the cube of interest is updated by summing the contribution of the sinogram elements. An advanced method of back projection is called filtered back-projection (FBP). This reconstruction technique reduces the amplitude of the low frequency components by filtering out each row of the sinogram [2]. The back-projection based reconstruction is mainly employed in Parallel Beam and Fan Beam SPECT systems. The back-projection based reconstruction is mathematically described in:

$$f(x,y) = \int_0^\pi g(s,\theta)d(\theta) \quad (2)$$

where, $f(x,y)$ is the reconstructed voxel by integrating the projection pixels $g(s,\theta)$ after filtering the sinogram by row for all view angles. Where $s$ and $\theta$ are detector location and view angle, respectively.

The main iterative algorithms are the Algebraic Reconstruction Technique (ART), the Maximum Likelihood Expectation Maximization (MLEM), the Ordered Subset Expectation Maximization (OSEM) and the Maximum A Posterori (MAP). These methods iteratively estimate the value $f(x,y)$ of the radionuclide distribution by assigning initial values to the voxels. Normally, all voxel values are initialized to 1.

In ART, the value of $f$ at $(x,y)$ is updated at every iteration based on the recursive equation [1]:

$$f_j^{k+1} = f_j^k + \frac{g_i - \sum_{j=1}^{N} f_{ji}^k}{N} \qquad (3)$$

where $f_k$ is the current value and $f_{k+1}$ is the new estimate. $g_i$ is the value of the projection value due to the $i^{th}$ ray. Moreover, $N$ is the total number of voxels that contribute to the projection pixel $g_i$. The new estimate, $f_j^{k+1}$ will be the sum of previous value and correction term. The correction term is the normalized error between the actual projection and the projection calculated from the estimated distribution.

MLEM and MAP are based on probabilistic machine learning models. The parametric model in Equation 1 is reformulated into a probabilistic model for MAP and MLEM, as in the following equation:

$$g|f \sim Normal(Af, \sigma) \qquad (4)$$

Equation 4 infers that the projection for a given reconstruction will be normally distributed with a mean of $Af$ and a standard deviation of $\sigma$. The variables $A$ and $\sigma$ are the parameters of the system. In SPECT reconstruction, instead of calculating the system matrix, we have programmed a forward projection model by using ray tracing.

The MLEM algorithm estimates a reconstructed object that can generate a likelihood projection to the actual projection [3]. This algorithm is implemented by approximating the distribution type of the radiation source as a Poisson's distribution. MLEM is composed of two steps: the expectation step and the maximization step. In the expectation step, the algorithm calculates any reconstructed image that provides a similar projection. The estimation of the reconstructed image with a good likelihood is calculated in the maximization step. The recursive formula is:

$$\bar{f}_j^{k+1} = \frac{\bar{f}_j^k}{\sum_{i=1}^{n} a_{ij}} \sum_{i=1}^{n} \frac{g_i}{\sum_{j'=1}^{m} a_{ij'} \bar{f}_{j'}^k} a_{ij} \qquad (5)$$

OSEM is similar to MLEM. However, in this method, projections are grouped into ordered subsets and are executed the same way as in MLEM [4]. This algorithm is introduced to improve the convergence of the estimation.

MAP aims to find the likelihood of the reconstructed image that gives the same projection while trying to reduce the noise in the reconstructed image [5]. The additional feature of noise reduction is introduced in MAP by adding a regularization term R, with a weighting $\beta$, as in:

$$\bar{f}_j^{k+1} = \frac{\bar{f}_j^k}{\sum_{i=1}^{n} a_{ij} + \beta R} \sum_{i=1}^{n} \frac{g_i}{\sum_{j'=1}^{m} a_{ij'} \bar{f}_{j'}^k} a_{ij} \qquad (6)$$

$\beta$ is the smoothing parameter and it should be low enough to avoid negative values in the denominator. The regularization term R is proportional to the derivative of the energy function, calculated as in:

$$R = \frac{\partial}{\partial f_j} U\left(\bar{f}_j^k\right) \qquad (7)$$

where $U$ is the energy of the system. In the above methods, despite the fast algorithm processing in case of FBP, the obtained resolution wouldn't be accurate due to smoothing filter. Whereas, MLEM estimate the scanned object iteratively. However, as the number of iterations increases, the noise in the reconstructed phantom rises. The estimated noisy reconstructions at different iterations can also provide likelihood projections. Moreover, the speed of convergence is also lower in MLEM compared to other methods. The speed can be improved using OSEM, with the same drawbacks of MLEM. MAP is introduced to include the likelihood of the radiotracer distribution for a given projection and to reduce noise.

These iterative reconstruction methods estimate a single reconstructed phantom that can generate a likelihood of the projection to the actual one. This implies that voxel values are deterministic and that the above methods don't consider any uncertainty in the reconstructed images.

By exploring the literature, several research applied Bayesian based uncertainty analysis in medical imaging for activity and attenuation estimation as presented by Leynes *et al.* [6], the authors applied Bayesian deep learning to estimate the uncertainty while generating a pseudo-CT from MRI data. This method is not applied for SPECT image reconstruction. Fuin *et al.* [7] tried estimating the uncertainty in SPECT reconstruction by using the Fisher Information Matrix (FIM). The FIM is used to calculate the uncertainty in terms of the variance of the reconstructed voxel values analytically. Moreover, this work calculates the uncertainties in the reconstruction by using the MAP estimation in terms of the FIM matrix.

The probabilistic programming concept can provide multiple values of each reconstructed voxel, which gives likelihood results of the projection with a specific distribution. This means that probabilistic programming considers the uncertainty in the reconstruction based on a particular distribution. The different reconstructed images within a certain uncertainty level can provide the likelihood of the actual position. The uncertainty calculation makes it possible to represent the reconstructed voxel value as a distribution instead of a fixed value. Normally, the reconstructed image is represented as $R = r_1, r_2, r_3 \ldots r_M$, where $M$ is the total number of voxels in the reconstructed phantom. The probabilistic programming identifies the uncertainty in the reconstruction as a normal distribution of each with means $\mu_1$, $\mu_2 \ldots \mu_M$ and variances $\sigma_1, \sigma_2, \ldots, \sigma_M$. Therefore, the reconstruction by using probabilistic programming can be represented as:

$$R_p = \{Normal(\mu_1, \sigma_1), Normal(\mu_2, \sigma_2), \ldots, Normal(\mu_M, \sigma_M)\} \quad (8)$$

This work focuses on generating uncertainties in the reconstruction for parallel hole SPECT by using the probabilistic programming concept. The details about the probabilistic programming reconstruction (PPR) algorithm are presented in section 2 with the technical details of Markov Chain Monte Carlo (MCMC) sampling and the forward projection implementation. The experiments and results are discussed in Section 3. Finally, conclusions are in section 4.

## 2. Probabilistic Programming for SPECT Reconstruction

Initially, the patterns of reconstructed images is defined using probability distributions. The probabilistic programming is based on the Bayesian inference [8]. The posterior value from the prior and likelihood of observations is calculated by:

$$P(f|g) = \frac{P(g|f).P(f)}{P(g)} \quad (9)$$

where $P(f|g)$ is the posterior distribution, $P(f)$ is the prior distribution, and $P(g|f)$ is the observation. In terms of image reconstruction, the posterior distribution is the conditional probability of the reconstruction for a given projection. Moreover, the observation is the conditional probability of the projection for a given reconstruction. The prior will be the probability of the reconstruction and $P(g)$ is the probability of the projection. Therefore, the conditional probability of the voxel values given their projection values will be proportional to the product of the likelihood value of the projection for a given reconstruction and the prior probability of reconstruction. The posterior value of the reconstruction is updated for every iteration. The probability of the projection is the integral of the joint probability of the projection and reconstruction concerning the change in the projection probability. However, the calculation of $P(g)$ becomes computationally complex as the dimension of the reconstruction is increased [9], because it must consider the contribution of each voxel value while determining the projections.

### 2.1 Approximate Bayesian inference

The computational complexity to calculate $P(g)$ is overcome by using an approximate Bayesian inference. In this method, it is possible to sample the posterior distribution of the reconstruction with respect to the projection by using joint probability. Therefore, it is not necessary to calculate the solution for the posterior probability by using a closed-form Bayesian method. Then the Bayesian equation for reconstruction will be updated as:

$$P(f|g) \propto P(g|f).P(f) \quad (10)$$

This posterior probability is generated by using MCMC sampling. These samplers output different posterior values of reconstruction with certain distributions that give likelihood values of projection. MCMC ensures that the sample we get at the end will be similar to the independent samples from the exact posterior distribution. There are different types of MCMC sampling available. In these algorithms, the transition from the current sample is carried out by proposing a new sample using some tuning parameters. The newly proposed sample will be considered only if the proposal is accepted with the known probability, otherwise it will be rejected. Then the next sample will be the same as the previous sample. In this work, we used the No-U-Turn Sampler (NUTS).

### 2.2 NUTS sampler

The NUTS sampling is an advanced version of the Hamiltonian Monte Carlo Sampling (HMC). The HMC is an MCMC sampling method that was developed to avoid random walk behaviour in other sampling methods [10]. In HMC, model samples are generated by simulating the behaviour of a physical system. In this system, particles in multidimensional space are subjected to potential and kinetic energies. So, the energies of these particles at specific energy levels can be defined using potential and kinetic energies, or in terms of position and momentum. The Hamiltonian term is the sum of the potential and kinetic energies. The Leap-Frog algorithm is used to find the position and momentum of the next sample during the HMC sampling. In this algorithm, the important step is to calculate the gradient of the potential energy, which is with respect to the position of the particle. The chain rule is used to calculate the gradient. It is possible to use forward differentiation and reverse differentiation to calculate the derivatives by using the chain rule. However, in the reconstruction context, the number of parameters is very high, so even though the reverse differentiation is fast, it takes a vast amount of memory [11]. The chain rule for reverse mode automatic differentiation is performed using [12]:

$$\frac{\partial y}{\partial x} = \frac{\partial y}{\partial w_1}\frac{\partial w_1}{\partial x} = \left(\frac{\partial y}{\partial w_2}\frac{\partial w_2}{\partial w_1}\right)\frac{\partial w_1}{\partial x} = \left(\left(\frac{\partial y}{\partial w_3}\frac{\partial w_3}{\partial w_2}\right)\frac{\partial w_2}{\partial w_1}\right)\frac{\partial w_1}{\partial x} = \cdots \quad (11)$$

In the reverse mode differentiation, the derivative of the dependent variable y is calculated with respect to its subexpressions $w_1$, $w_2$, etc. This derivative calculation helps the HMC to converge quickly. However, the HMC mainly depends on two parameters, namely the step size and the discretization time. If the step size is too small, the HMC starts exhibiting random walk behaviour, and if it is too high, it takes too long to reach convergence. Thus, the HMC requires some tuning. However, the NUTS uses a recursive algorithm to generate a set of likely sample points with a given distribution. Moreover, the sampler automatically stops when it starts to double back and retrace its steps [13]. The NUTS is implemented with algorithms to autotune the step size and the discretization time during sampling.

Apart from this, all samples are drawn with a specific distribution. So, prior knowledge of the system is beneficial to find the best reconstructions with the highest likelihood. If the selected distribution is normal, then specifying the mean and variance of the distribution is essential. In model-based reconstruction, we need to specify the observations. In medical imaging reconstruction, projections are the observations. These observations can be specified with a normal distribution with mean value as the projection, calculated by using the prior reconstructions and a standard deviation as a constant value.

## 2.3 Algorithm for probabilistic programming based reconstruction

The pseudo algorithm for probabilistic programming based reconstruction consists of the following steps:
1. Define the probabilistic model.
   a. Assign distribution to voxel values.
   b. Find the forward projection based on the voxels' values with the assigned distribution.
   c. Assign distribution to the projection by setting forward projection value from step b as mean and a fixed variance.
2. Sampling using a specified sampling algorithm.

The prior of the reconstruction can be a distribution of any type. The one substantial condition is that the values of the prior should be greater than zero. Similarly, the distribution of the observation, that is of the projection, is assigned based on the forward projection generated with the prior values of the reconstruction. The forward projection algorithm is explained in section 2.4. This model is used for sampling with the HMC algorithm to find the posterior distribution through different iterations.

## 2.4 Forward projection

The algorithm for forward projection is implemented based on the fast voxel ray traversal algorithm, tracking a ray through a cube. It outputs the co-ordinate of each voxel through which the ray has passed, and the entry and exit time in that voxel. Using the entry and exit time, we can calculate the intersection length inside that voxel. For the forward projection, first, the center of the detector pixel is calculated and a vertical ray towards the cube of interest is generated. Then, the ray is traced through the cube, and each voxel intersection is identified [14-15]. Then, the detector pixel location, from where the ray is generated, is updated with the sum of the intersected voxels. Repeating this step for all detector pixels. After that, the detector base is rotated one step angle and the above procedure was repeated. These projections for each view angle are vectorized and concatenated into a single matrix. This concatenated projection matrix is used for reconstruction.

## 2.5 Julia programming and Turing toolbox

The probabilistic programming based SPECT image reconstruction is implemented in the Julia programming platform with the help of the Turing Toolbox package [16]. This software is a high-performance computing language with similar performance as C programming and similar ease of use as the Python language. Moreover, the LLVM compiler used in this software package makes programs run faster compared to other languages. The Turing toolbox was originally developed for probabilistic programming and used for uncertainty generation. The Turing toolbox has built-in functions for different MCMC samplers. In this work, we use the NUTS sampler only. After the sampling, the output contains values of each reconstructed voxel at different sampling, with the mean of these reconstruction samples.

## 2.6 Distributed programming

Using the probabilistic method, the reconstruction time is found to increase quadratically with the number of voxels in the phantom. Increasing the number of iterations increases the likelihood of a more probable reconstruction. The distributed programming capability of the Julia programming language is incorporated into the probabilistic programming to speed up the whole process.

## 3. Results

The experiments were conducted using a NUTS sampler and reverse differentiation backend. Objects with 4x4x4, 8x8x8 and 16x16x16 voxels were studied to analyze the point source response. Moreover, the uncertainty in the reconstruction of these point sources is analyzed in terms of variance. The Shepp-Logan phantom was used for scanning and reconstruction using probabilistic programming.

## 3.1 Reconstruction for different cube sizes

The performance of the probabilistic programming based reconstruction algorithm is analyzed with phantoms of size 4x4x4, 8x8x8, and 16x16x16. The forward projections are acquired based on the algorithm mentioned in section 2.4.

## 3.2 Point source reconstruction

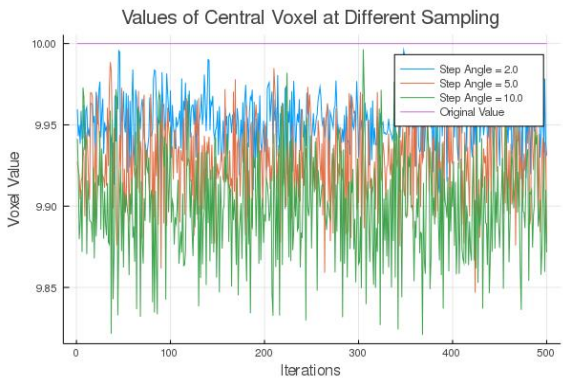

Figure 1: Sampling values of central voxel of phantom with 8x8x8 voxels at different step angles

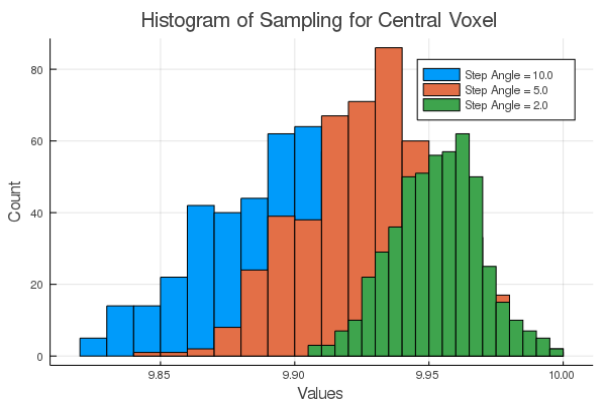

Figure 2: Histogram of generated samples for central voxel for different view angles

A point source reconstruction is performed using a phantom with the central voxel with not-zero value and all other voxels with zero value. Projection images are calculated for various step angles, 2, 5 and 10 degrees. The scanned object is reconstructed probabilistically after many iterations. The mean value of the reconstructed phantoms with uncertainties is found by calculating the mean of all samples after iteration. The uncertainties pertaining to the central voxel after reconstruction are shown in Figure 1. These uncertainities in reconstruction can generate projections with values similar to actual scanned values. The sampling values at different iterations for step angles 2.0 degree, 5.0 degree and 10.0 degree are drawn in blue, red and green colours respectively. The pink represents the actual phantom value. Figure 1 illustrates that the sampling values converge closer to the original value when the number of projections increases.

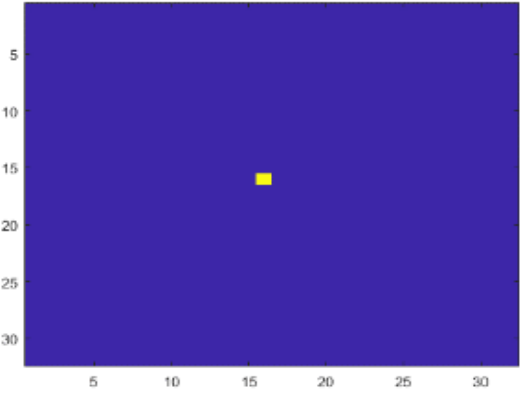

Figure 3: Mean reconstructed images having voxels 32x32x32

*3.2.1 Histogram of sampling:* The histograms provide information about the number of data points that are generated by a probabilistic fall within a range of values of each voxel. These data points are generated as a result of probabilistic programming based reconstruction. The histogram is plotted after taking samples of each voxel after a specified number of iterations. The mean value of the uncertainty is the peak of the histogram. The histogram of reconstructions with different step angles is shown in Figure 2. The reconstruction corresponds to the histogram, whose peak is closest to the actual voxel value generates best likely projections. It is evident from Figure 2 that the histogram comes close to the actual value if the step angle is small.

*3.2.2 Point spread analysis:* Point spread analysis is the impulse response analysis of an optical system. In this analysis, the central voxel value of the phantom is set to a value higher than zero. Here for the analysis, the central voxel value and the remaining voxel values are selected as 10 and zero, respectively. The projection of the phantom is generated for each view angle using ray tracing, and it is used for the reconstruction. The mean reconstructed image calculated from all samples is used for point spread analysis. The central slice of the mean reconstructed image is shown in Figure 3.

### 3.2.3 Full Width at half maximum (FWHM) analysis:

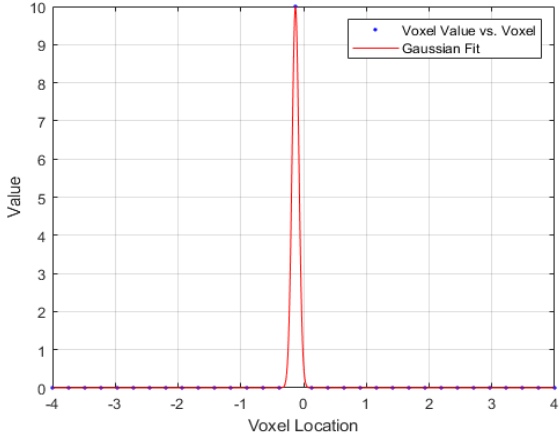

Figure 4: Gaussian fitted point source response of reconstructed image with 32x32x32 voxels.

Table 1: FWHM of reconstructed objects with 32x32x32 voxels by using projections with different step angles

| Step Angle (degree) | 2 | 5 | 10 |
|---|---|---|---|
| FWHM (mm) | 0.2581 | 0.2582 | 0.2582 |

FWHM is used for the spatial resolution of an optical system. The point source reconstructed image is used for FWHM analysis. Initially, the mean reconstructed image is calculated using samples from the probabilistic reconstruction. The central row of the central slice is extracted from the mean reconstructed image is plotted against the row location. This plot is fitted to a Gaussian curve as shown in Figure 4. The width of this gaussian fitted curve at half of the maximum gives the FWHM value. The FWHM of the point source responses at different iterations is given in Table 1.

### 3.3 Comparison between MLEM and MAP

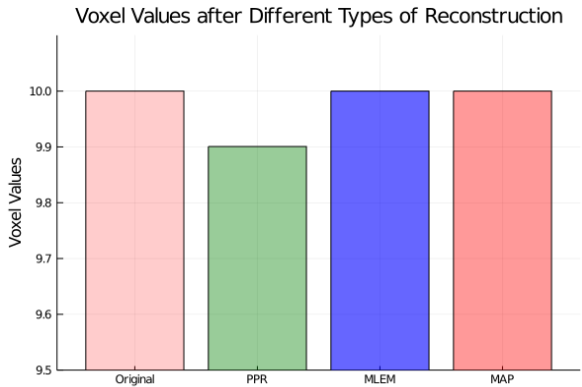

Figure 5: Central voxel value comparison by using different types of reconstruction algorithm

MLEM and MAP are the two main iterative reconstruction techniques. These algorithms generate optimized values of the reconstructed phantom. However, these algorithms do not provide any uncertainty information. The mean reconstructed image generated using probabilistic programming is compared with the one reconstructed by using MLEM and MAP. The comparison analysis with MLEM and MAP is conducted both visually and analytically. The reconstructed images in Figure 5 show how close the probabilistic reconstruction is to conventional reconstruction methods. The relative norm is used for the comparison of the reconstruction quality. The ideal value of relative norm is zero for perfect reconstruction. However in reality, if the relative norm is small, the voxel values of the reconstructed phantom will be closer to the activity in the scanned phantom. The relative norm, $R_{norm}$ is calculated as:

$$R_{norm} = \frac{\sqrt{(x_1-\overline{x_1})^2+(x_2-\overline{x_2})^2+(x_3-\overline{x_3})^2....(x_n-\overline{x_n})^2}}{\sqrt{(x_1)^2+(x_2)^2+(x_3)^2....(x_n)^2}} \quad (12)$$

Where, $x_1, x_2, x_3 ... x_n$ are the actual phantom voxel values and $\{\overline{x_1}, \overline{x_1}, \overline{x_2}, \overline{x_3}, ....\overline{x_n}\}$ are the reconstructed phantom values.

Table 2: Relative norm of reconstructions by applying different algorithms

| Number of Voxels | Relative Norm | | |
|---|---|---|---|
| | MAP | MLEM | PPR |
| 4x4x4 | 0.00557 | 0.00557 | 0.009 |
| 8x8x8 | 0.0304 | 0.0304 | 0.01215 |
| 16x16x16 | 0.0827 | 0.155 | 0.025607 |

The relative norms of reconstruction using probabilistic and conventional algorithms including MAP and MLEM are discussed in Table 2. This table shows that the norm of the mean reconstructed image using probabilistic programming is nearly equal to the norm of MLEM and MAP based reconstruction. This means that the accuracy of reconstruction using a probabilistic programming method is closer to MLEM and MAP based reconstruction

### 3.4 Uncertainty analysis

The uncertainty in the reconstruction is calculated in terms of variance, mean of variance and histogram of variance. Initially, the variance of each voxel is calculated from samples obtained using probabilistic programming. If the variances of voxels are very low, then the forward projection model used in the reconstruction is close to the actual model and reconstruction values will be close to the actual phantom values. If the variance is too high, this will be a clear indication of the inaccuracy in the forward projection and reconstruction model. Moreover, the chance of error in the reconstruction can be higher for high variance cases.

Table 3: The variance of central voxel after reconstructed by using probabilistic programming

| Step Angle (in degree) | 2.0 | 5.0 | 10.0 |
|---|---|---|---|
| Number of Voxels | Variance of central Voxel | | |
| 4x4x4 | 0.0225 | 0.0225 | 0.0225 |
| 8x8x8 | 0.0162 | 0.0252 | 0.0325 |
| 16x16x16 | 0.0176 | 0.0265 | 0.0321 |

Table 3: The mean of variances of phantoms of different sizes scanned at different view angles

| Step Angle (in degree) | 2.0 | 5.0 | 10.0 |
|---|---|---|---|
| Number of Voxels | Mean of Variances | | |
| 4x4x4 | 0.008 | 0.008 | 0.008 |
| 8x8x8 | 0.004 | 0.006 | 0.009 |
| 16x16x16 | 0.003 | 0.005 | 0.006 |

*3.4.1 Variance of each voxels at different iterations:* In this work, the effect of variance between samples of each voxel are studied and analyzed for different step angles. No significant effect of view angles in the variance is found when the number of voxels is low. However, the effect of step angle in variance, calculated from the samples generated using probabilistic reconstruction, is significant. It was found to decrease for small step angles. The variance of the central voxel at different numbers of iterations and different step angles is shown in table 3.

*3.4.2 Mean of variances:* The mean of variance calculation is also a way to express the uncertainty in probabilistic programming based reconstruction. Initially, variances in the reconstruction are generated by using the method described in section 3.4.1. After that, the mean of variances of all voxel values is calculated using simple statistical methods. The mean of variances of different specifications is discussed in Table 4.

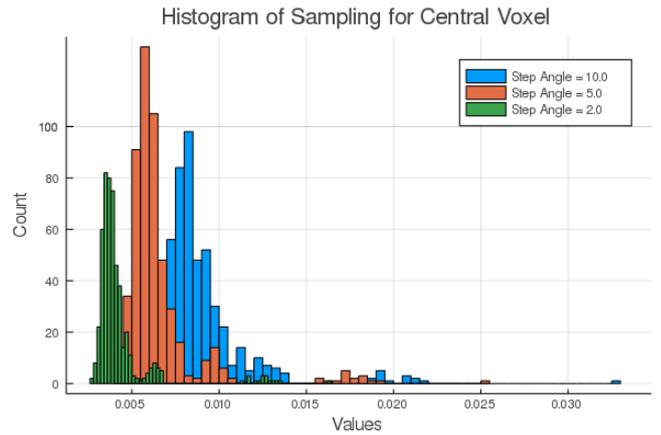

Figure 6: Histogram of variances reconstructions after taking projections at different step angles

*3.4.3 Histogram of variances:* This histogram of variance is also used to visualize uncertainty in reconstruction. The variance of each sample is calculated by using a simple statistical technique. Then the histogram of voxels variance is plotted. Figure 6 shows the histogram for different SPECT settings. This histogram indicates that the variance approaches the value zero, when the step angle approaches the zero value. This indicates that uncertainty reduces when the step angle reduces.

*3.5 Shepp Logan phantom reconstruction*

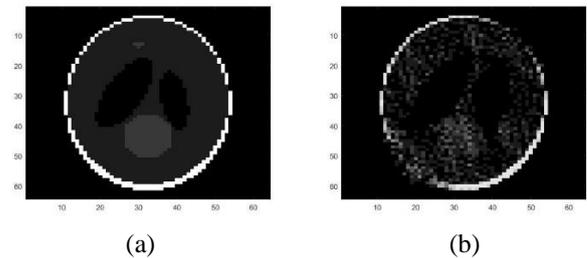

(a)                 (b)

Figure 7: Central slice of (a) Original Shepp-Logan phantom, (b) Mean reconstructed image after probabilistic programming based reconstruction.

The Shepp-Logan phantom is a standard phantom used for medical image reconstruction simulations. It is considered as a standard phantom in medical image reconstruction studies. The projection of the Shepp Logan phantom is calculated using the projection algorithm discussed in section 2.4. Further, these projections of the phantoms were used as input for probabilistic reconstruction and analysis. The samples were generated using probabilistic reconstruction and they represent the uncertainties in the reconstruction. The mean reconstructed images after probabilistic reconstruction is used for visual comparison with original phantom and the central slices are shown in figure 7.

## 4. Conclusion

The forward projection model has a substantial effect on iterative reconstruction quality. The error in this model may lead to an inaccurate reconstruction of the scanned object. The probabilistic programming methods sampled these uncertainties in reconstruction by using prior knowledge of the system. In this work, the No-U-Turn algorithm based sampling mechanism was employed for generating uncertainties. These samples for uncertainty calculation were able to generate likely scanned images to the actual one. Experiments generated uncertainties as variance, and the histogram of variance exhibited gaussian nature. Moreover, the variance and the mean of variance approached to zero value when the difference between adjacent scanning angle decreases. The scanned object was reconstructed from the uncertainties by taking the mean of uncertainties, and the precision was closer to MAP and MLEM based reconstructions. However, probabilistic programming-based reconstruction is a time-consuming process if the number of voxels is higher in the reconstructed object. This time consumption can be improved by using distributed programming or by optimizing the sampling algorithms..

## References

[1] Bruyant, P. P. (2002). Analytic and iterative reconstruction algorithms in SPECT. *Journal of Nuclear Medicine, 43*(10), 1343-1358.

[2] Tsui, B. M. W., & Frey, E. C. (2006). Analytic Image Reconstruction Methods in Emission Computed Tomography. In H. Zaidi (Ed.), *Quantitative Analysis in Nuclear Medicine Imaging* (pp. 82-106). Boston, MA: Springer US.

[3] Shepp, L. A., & Vardi, Y. (1982). Maximum likelihood reconstruction for emission tomography. *IEEE Trans Med Imaging, 1*(2), 113-122.

[4] Hudson, H. M., & Larkin, R. S. (1994). Accelerated Image-Reconstruction Using Ordered Subsets of Projection Data. *Ieee Transactions on Medical Imaging, 13*(4), 601-609.

[5] Green, P. J. (1990). Bayesian Reconstructions from Emission Tomography Data Using a Modified Em Algorithm. *Ieee Transactions on Medical Imaging, 9*(1), 84-93.

[6] Leynes, A. P., Ahn, S. P., Wangerin, K. A., Kaushik, S. S., Wiesinger, F., Hope, T. A., & Larson, P. E. (2020). Bayesian deep learning Uncertainty estimation and pseudo-CT prior for robust Maximum Likelihood estimation of Activity and Attenuation (UpCT-MLAA) in the presence of metal implants for simultaneous PET/MRI in the pelvis. *arXiv preprint arXiv:2001.03414*.

[7] Fuin, N., Pedemonte, S., Arridge, S., Ourselin, S., & Hutton, B. F. (2014). Efficient Determination of the Uncertainty for the Optimization of SPECT System Design: A Subsampled Fisher Information Matrix. *Ieee Transactions on Medical Imaging, 33*(3), 618-635.

[8] Gamerman, D., & Lopes, H. F. (2006). Markov chain Monte Carlo: stochastic simulation for Bayesian inference: CRC Press.

[9] Bertorelle, G., Benazzo, A., & Mona, S. (2010). ABC as a flexible framework to estimate demography over space and time: some cons, many pros. *Molecular ecology*, 19(13), 2609-2625.

[10] Neal, R. M. (2011). MCMC Using Hamiltonian Dynamics. *Handbook of Markov Chain Monte Carlo*, 113-162.

[11] Christianson, D. B., Davies, A. J., Dixon, L. C. W., Roy, R., & VanderZee, P. (1997). Giving reverse differentiation a helping hand. *Optimization Methods & Software, 8*(1), 53-67.

[12] Baydin, A. G., Pearlmutter, B. A., Radul, A. A., & Siskind, J. M. (2017). Automatic differentiation in machine learning: a survey. The Journal of Machine Learning Research, 18(1), 5595-5637.

[13] Hoffman, M. D., & Gelman, A. (2014). The No-U-Turn Sampler: Adaptively Setting Path Lengths in Hamiltonian Monte Carlo. *Journal of Machine Learning Research, 15*, 1593-1623.

[14] Amanatides, J., & Woo, A. (1987). *A fast voxel traversal algorithm for ray tracing.* Paper presented at the Eurographics.

[15] Siddon, R. L. (1985). Fast calculation of the exact radiological path for a three-dimensional CT array. *Medical Physics, 12*(2), 252-255.

[16] Ge, H., Xu, K., & Ghahramani, Z. (2018). Turing: a language for flexible probabilistic inference. *International Conference on Artificial Intelligence and Statistics, Vol 84, 84*.